\documentclass[conference]{basi}
\usepackage[british]{babel}
\usepackage[varg]{txfonts}
\usepackage{rotating}
\usepackage{dcolumn}
\begin{document}
\title[Young eruptive low mass stars]{The outburst and nature of young 
eruptive low mass stars in dark clouds}
\author[J.~P.~Ninan et~al.]%
       {J.~P.~Ninan$^1$,
       D.~K.~Ojha$^1$\thanks{email: \texttt{ojha@tifr.res.in}}, B.~C.~Bhatt$^2$, 
K.~K.~Mallick$^1$, A.~Tej$^3$, 
\newauthor D.~K.~Sahu$^2$, S.~K.~Ghosh$^1$ and V.~Mohan$^4$\\
       $^1$Tata Institute of Fundamental Research, Homi Bhabha Road, Colaba,
Mumbai 400 005, India\\
       $^2$CREST, Indian Institute of Astrophysics, Koramangala, Bangalore 560 034, India\\
$^3$Indian Institute of Space Science and Technology, Valiamala P.O., \\Thiruvananthapuram 695 547, India\\
$^4$Inter-University Centre for Astronomy and Astrophysics, Post Bag 4, Ganeshkhind, \\Pune 411 007, India
}



\date{}
\maketitle
\label{firstpage}

\begin{abstract}
The FU Orionis (FUor) or EX Orionis (EXor) phenomenon has attracted 
increasing attention in recent years and is now accepted as a crucial 
element in the early evolution of low-mass stars. FUor and EXor eruptions of 
young stellar objects (YSOs) are caused by strongly enhanced accretion from 
the surrounding disk. FUors display optical outbursts of $\sim$ 4 mag or more and 
last for several decades, whereas EXors show smaller outbursts 
($\Delta$m $\sim$ 2 - 3 mag) that last from a few months to a few years and 
may occur repeatedly. Therefore, FUor/EXor eruptions represent a rare but 
very important phenomenon in early stellar evolution, during which a young 
low-mass YSO brightens by up to several optical magnitudes. Hence, long-term 
observations of this class of eruptive variable are important to design 
theoretical models of low-mass star formation. In this paper, we present 
recent results from our long-term monitoring observations of three rare types 
of eruptive young variables with the 2-m Himalayan {\it Chandra} Telescope (HCT) 
and the 2-m IUCAA Girawali Observatory (IGO) telescope.
\end{abstract}

\begin{keywords}
stars: formation -- stars: pre-main-sequence -- stars: variables: other --
ISM: clouds -- (ISM:) reflection nebulae
\end{keywords}

\section{Introduction}

There is now convincing evidence that EXor and FUor outburst phenomena 
are closely related to the earliest
stages of stellar evolution (Herbig, Petrov \& Duemmler 2003). 
Rarity and
obscuration have resulted in a poor understanding of the eruption
mechanism in spite of an established integral link to disk accretion
(Reipurth \& Aspin 2004; Hartmann, Hinkle \& Calvet 2004).
FUor and EXor,
also referred to as sub-FUors, are among the most interesting and intriguing
class of known pre-main-sequence (PMS) stars. They are likely to be near
solar-mass protostars that are still accreting material from their 
circumstellar disks and are associated
with collimated outflows (Sandell \& Weintraub 2001). Only a handful
of these objects are known to date (e.g., Vittone \& Errico 2005).
Hence, there is an urgent need for observations and studies 
to facilitate a deeper understanding of their nature and the 
effects of their associated eruptions. The morphology and nature of small compact
reflection nebulae in the star-forming clouds possibly hint at these objects 
being in a transition phase between an embedded PMS star and a visible Herbig-Haro
object (Reipurth \& Bally 2001; Reipurth \& Aspin 2004). A similar
case was the emergence of McNeil's nebula, which was found
to harbour a possible EXor event (Ojha et al. 2006 and references
therein).

In this paper, we present optical observations of the post-outburst
phases of McNeil's nebula (V1647 Ori), a new reflection nebula in LDN 1415, 
and a cometary reflection nebula associated with the infrared source 
IRAS 06068-0641. 
In Section 2, we present details of the observations and
data reduction procedures and in Section 3 we present the results and discuss
the short- and long-term variability of three eruptive
low-mass young variables.

\section{Observations and Data Reduction}

Optical ($V, R, I$) photometric observations were carried out on
several nights 
with the Hanle Faint Object Spectrograph Camera (HFOSC) and
IUCAA Faint Object Spectrograph Camera (IFOSC) on the 2-m
HCT, Hanle and the 2-m IGO telescope, Pune, respectively, 
during the period February 2004 - March 2011.
The HFOSC instrument, equipped with a SITe 2K $\times$ 4K
pixel CCD and the IFOSC instrument, equipped with an 
EEV 2K $\times$ 2K pixel thinned, back-illuminated CCD were used.
The field-of-view (FOV) for HFOSC, where only the central 2K $\times$ 2K region is
used for imaging, is  $\sim$ 10 $\times$ 10 arcmin$^2$
with a pixel scale of 0.296 arcsec and that of IFOSC is 
$\sim$ 10.5 $\times$ 10.5 arcmin$^2$
with a pixel scale of 0.307 arcsec. Photometric standard stars
(Landolt 1992) were observed on several nights to obtain
the atmospheric extinction and transformation coefficients.
For nights when the Landolt standard stars
were not available, we calibrated the data using
secondary standards present in the image frames. The average seeing 
(full width at half-maximum) in all bands was $\sim$ 1.8 arcsec
and 1.3 arcsec during our HCT and IGO observations, respectively.

Data reduction was done using the National Optical Astronomy Observatories'
(NOAO) IRAF\footnote{IRAF is distributed by the NOAO, which are operated by
the Association of Universities for Research in Astronomy, Inc., under
contract with the National Science Foundation.} package tasks. Object frames
were flat-fielded using the median combined normalized flat frames.
Identification and aperture photometry of point sources were performed
using the DAOFIND and DAOPHOT packages, respectively.
Given the nebulosity around V1647 Ori and IRAS 06068-0641, photometry was 
obtained using the point-spread function algorithm ALLSTAR in the DAOPHOT 
package (Stetson 1987). The residuals to the photometric solution 
are $\le$ 0.05 mag. 
Since the outburst source was too faint to be detected in the $VRI$ bands,
we used an aperture radius of 30 pixels ($\sim$ 10 arcsec) for photometry of
the L1415 nebula. This aperture radius was chosen to cover the entire
nebular emission and enable direct comparison with the results of 
Stecklum, Melnikov \& Meusinger (2007) who used the same aperture radius. 
The local sky was evaluated in an annulus with an inner radius
of 128 pixels and a width of 20 pixels. The residuals of the photometric
solution are $\le$ 0.04 mag. 

The color composite images of the central region of McNeil's nebula, 
the LDN 1415 nebula and the region around IRAS 06068-0641 were constructed from 
the HFOSC and IFOSC $V$, $R$ 
and $I$-band images ($V$ represented in blue, $R$ in green, and $I$ in red) 
and are shown in Figure 1.
 
\begin{figure}
\centerline{\includegraphics[width=4.3cm]{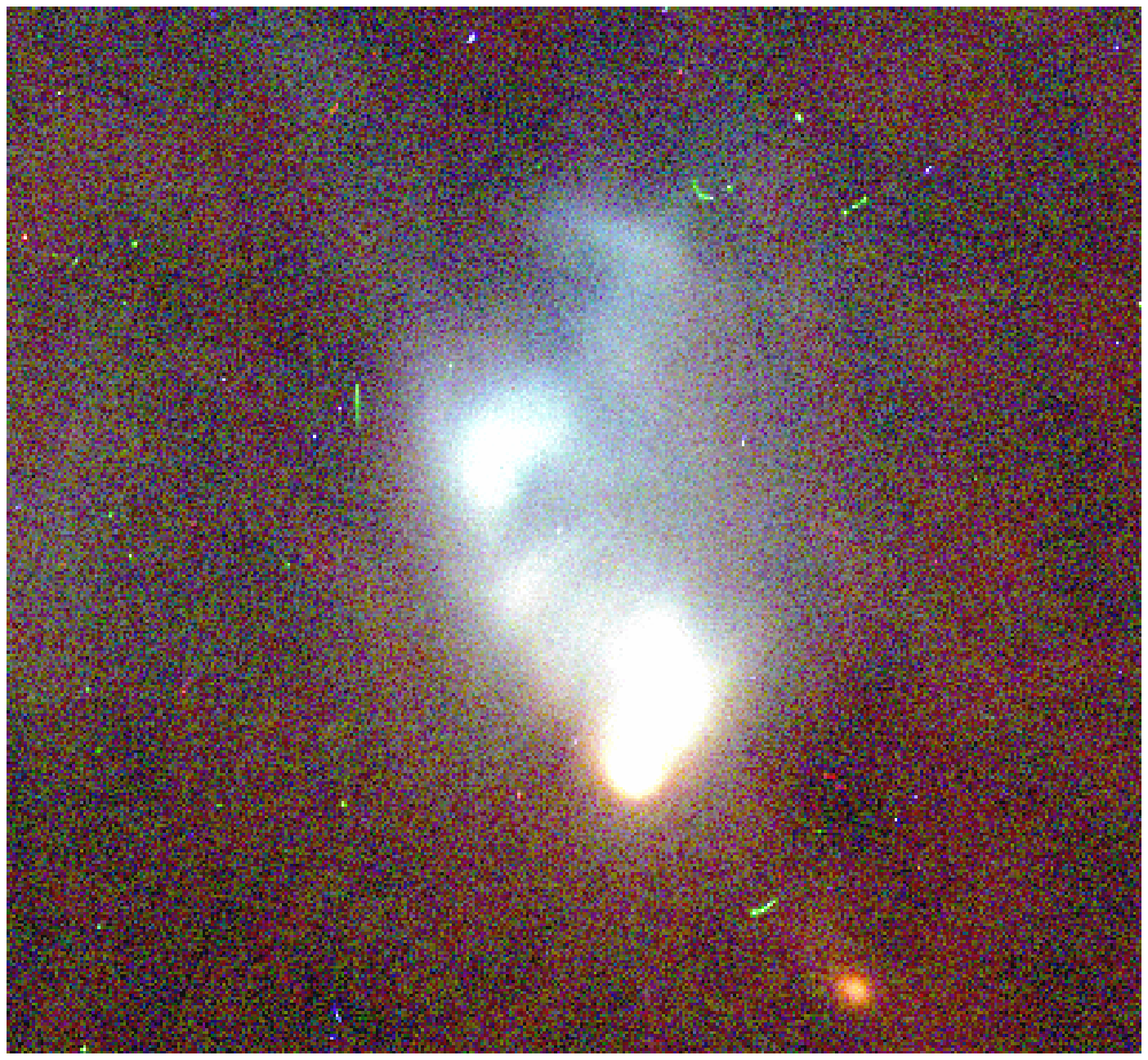} \qquad
            \includegraphics[width=4.05cm]{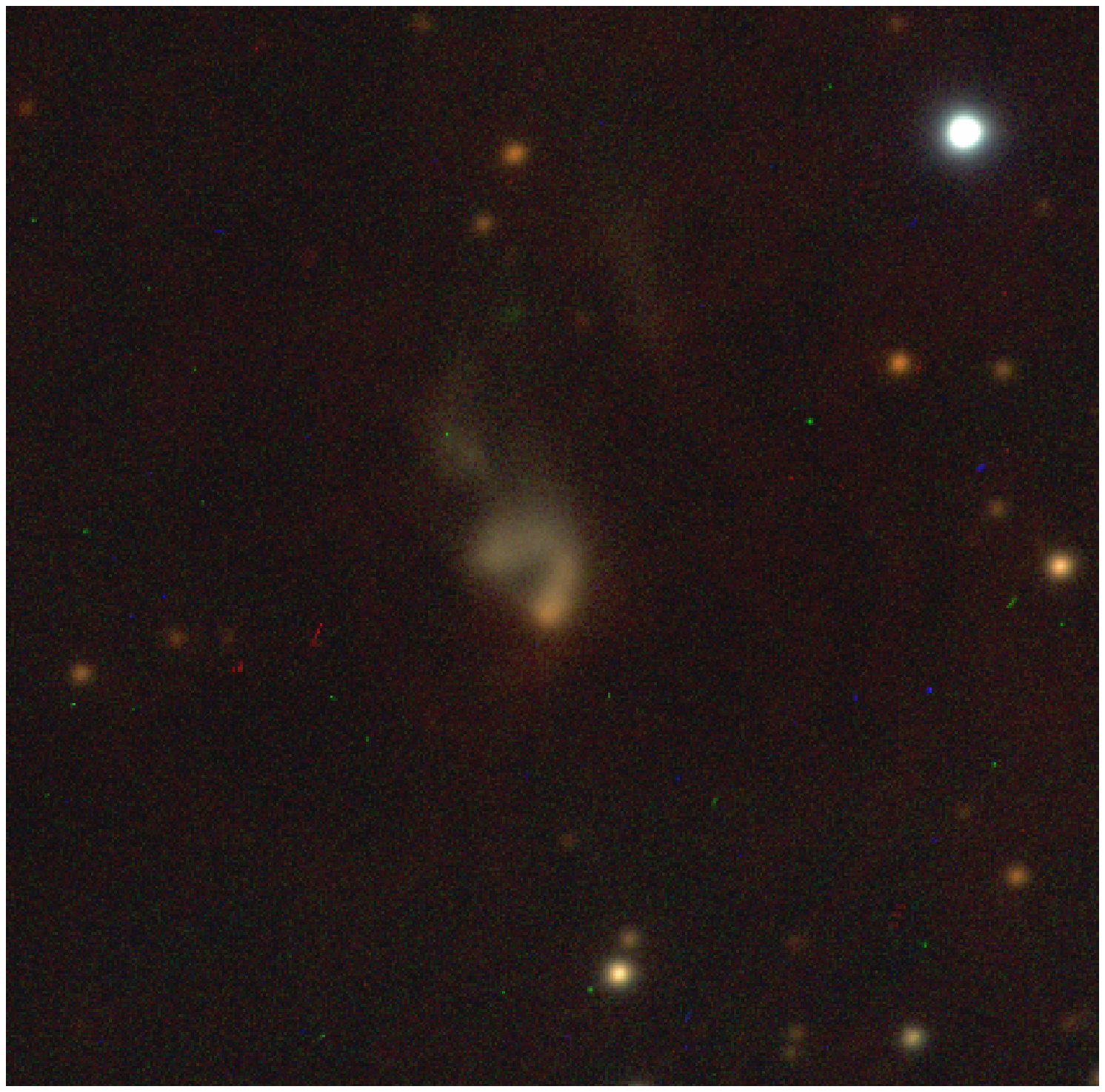} \qquad 
           \includegraphics[width=4cm]{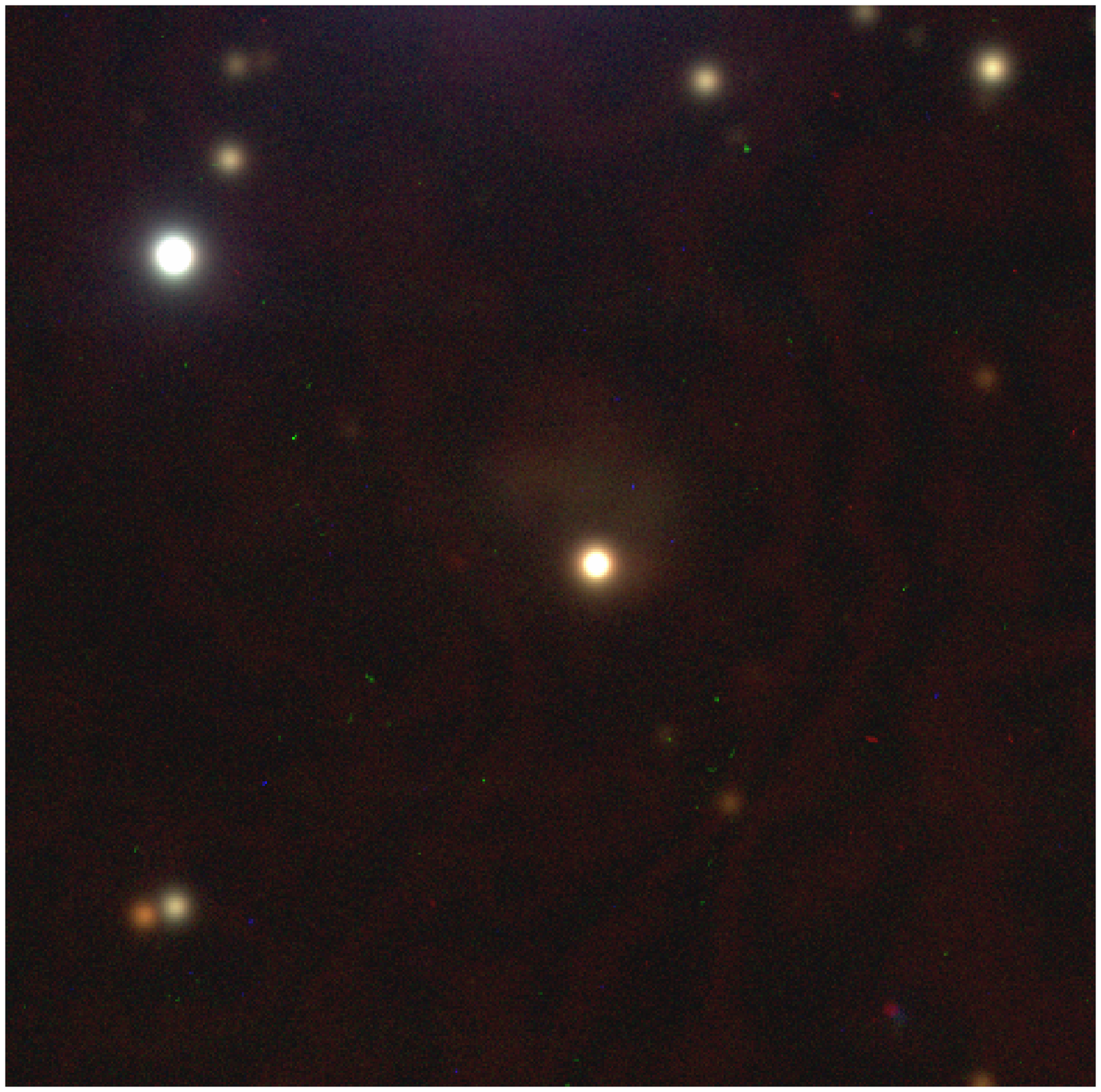}}
\caption{$VRI$ three-color composite images ($V$: blue, $R$: green, and $I$: red) 
of McNeil's nebula ({\it left}), the LDN 1415 nebula ({\it middle}) and 
the IRAS 06068-0641 region ({\it right}), obtained with HFOSC and IFOSC on
13 January 2011, 20 February 2011 and 2 January 2011, respectively. The outburst 
source of McNeil's nebula is located at the lower tip of the nebula.}
\end{figure}

\section{Results and Discussion}

\subsection{McNeil's nebula (V1647 Ori)}

The compact source at the base of a variable nebula (McNeil's Nebula Object) 
in the Lynds 1630 dark cloud in Orion went into outburst in late 2003
(McNeil 2004). Later, the McNeil's Nebula Object was identified as V1647 Ori 
(Samus 2004). V1647 Ori, a low-mass, deeply embedded, PMS star, 
has undergone two optical/near-infrared outbursts in the last decade, 
both of which gradually faded over several months to years. 
These eruptions are thought to have been the result of large-scale accretion events.

Ojha et al. (2006) presented a detailed study of the post-outburst phase
of McNeil's nebula using optical
($B,V,R,I$) and near-infrared ($J,H,K$) photometric and low-resolution optical
spectroscopic observations. The long-term optical and near-infrared observations
showed a general decline in the brightness of the exciting source of McNeil's
nebula, V1647 Ori. Our optical images taken in November 2005 showed that
V1647 Ori had faded by more than 3 magnitudes since February 2004 (see
Figure 2, {\it top panel}). McNeil's
nebula itself had also faded considerably. The optical spectra showed 
strong H$\alpha$ emission with blue-shifted absorption
and the Ca II IR triplet (\mbox{8498 \AA}, 8542 \AA~and 8662 \AA) in emission. The
presence of the Ca II IR triplet in emission confirmed that V1647 Ori is
a PMS star. Therefore, our long-term, post-outburst
photometric and spectroscopic observations of V1647 Ori indicated an EXor
rather than an FUor event.

\begin{figure}
\centerline{\includegraphics[width=11cm]{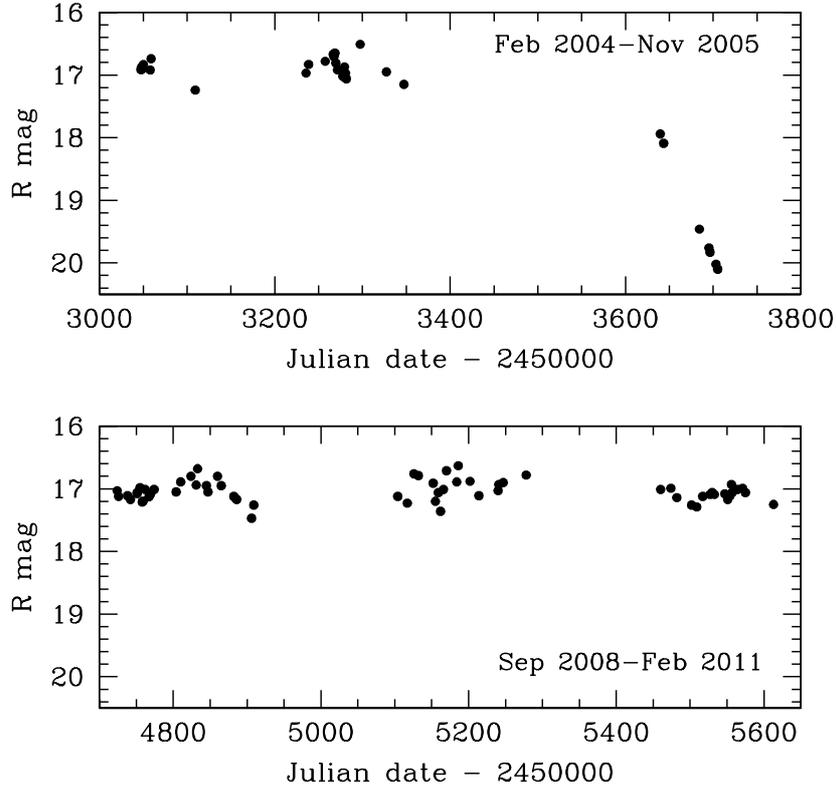}}
\caption{The optical light curve of V1647 Ori in $R$ band. The filled 
circles show HFOSC and IFOSC measurements from February 2004 to 
February 2011.}
\end{figure}

In 2008, V1647 Ori again underwent a strong outburst (Aspin 2008),
which is rarely seen in the early phases of low-mass young stellar objects.
We monitored V1647 Ori with the HCT and IGO telescopes beginning 
with the second outburst. No significant variation in
brightness of V1647 Ori was seen for about two months since 
the second outburst began (Ojha et al. 2008).
In comparison with the last reported quiescent phase (Ojha et al. 2006),
however, there was a brightening of about $\sim$ 3 magnitudes in $R$, and the
infrared colors suggested that circumstellar matter of A$_V$ $\sim$ 7.5 mag
had probably been cleared during this outburst. 
Figure 2 ({\it bottom panel}) shows the optical
light curve (September 2008 - February 2011) of V1647 Ori in $R$-band.
Comparison of the spatial flux distribution of the nebula with the first
post-outburst phase in 2004 revealed a change in the dust distribution
around the source during the second outburst. From our 2+ year long
monitoring observations (September 2008 - February 2011), we see significant
short-term variations in the brightness of V1647 Ori since the
second outburst began. The source, however, has not faded
away considerably as seen in 2004 - 2005. The source magnitude and 
\mbox{1-$\sigma$} fluctuations
in $V$, $R$ \& $I$ during the period of our observations were 18.86$\pm$0.23,
17.04$\pm$0.16, 14.99$\pm$0.12, respectively.

The observed properties of the outburst of V1647 Ori are
different in several respects from both the EXor and FUor type outbursts,
and suggest that this star represents a new type of eruptive
young star, one that is younger and more deeply embedded than EXor, and exhibits
variations on shorter time scales than FUors. 

\subsection{LDN 1415 (IRAS 04376+5413)}

The new reflection nebula in the not so well studied 
Lynds opacity class 3 dark cloud LDN 1415 (Lynds 1962) 
was first detected by Stecklum (2006) in early April 2006
in the vicinity of IRAS 04376+5413.
Stecklum, Melnikov \& Meusinge  (2007) later reported the 
presence of a new
compact arc-shaped nebula with a size of 20 arcsec in the CCD images of
the dark cloud LDN 1415. The brightness peak of the
nebula is within the positional error ellipse of IRAS 04376+5413.
Optical spectra of the nebula taken by Stecklum, Melnikov \& Meusinger 
on 21 September 2006 revealed the presence of a P-Cygni profile in the
H$\alpha$ line, indicating clear evidence for an FUor or EXor-type outburst
due to temporarily enhanced accretion. 
Kastner et al. (2006) observed this eruptive source with the
{\it Chandra} X-ray Observatory's Advanced CCD Imaging Spectrometer imaging array
(ACIS-I). No X-ray sources were detected, which constrained the X-ray
luminosity of the emergent source to be less than
$\sim 2 \times 10^{28}$ erg $s^{-1}$, assuming
the distance to the LDN 1415 cloud to be 170 pc.

To study the
post-outburst phase of the embedded source in the LDN 1415 nebula, we have
been carrying out optical observations of this source with 
HCT and IGO telescopes.
We present in Figure 3 variability measurements of the LDN 1415 nebula for a 
duration of about two and half years.
In comparison with the available pre-outburst photometry from POSS II
(epoch December 1996) and the KISO (January 2001) quoted in 
Stecklum, Melnikov \& Meusinger (2007), 
our first post-outburst data point shows an enhancement of $\sim$ 3.4 mags
in the $I$-band. Following this observation, a general decline in the 
brightness is seen
in all three ($VRI$) optical light curves (see Figure 3). 
Superimposed on this decline, we see
the presence of small-scale fluctuations of $\sim$ 0.2 - 0.3 mags over 
short time
scales of 3 - 8 months. This variation is consistent with the young and eruptive
nature of this class of objects. Therefore, our long-term, post-outburst 
optical and NIR photometric and optical
spectroscopic monitoring of the LDN 1415 nebula and its associated outburst source
from October 2006 to March 2009 (Pawade et al. 2010) suggest an EXor or FUor event,
possibly by the least luminous member of the known sample of FUor and EXor
objects (Stecklum, Melnikov \& Meusinger 2007).

\begin{figure}
\centerline{\includegraphics[width=11cm]{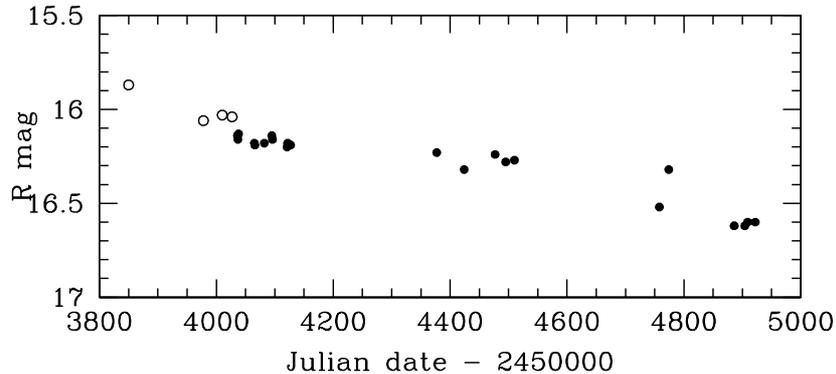}}
\vspace*{-5.5cm}
\caption{The optical light curve of L1415 nebula in $R$ band.
The filled circles show our HFOSC and IFOSC measurements
(October 2006 - March 2009). The empty circles show the photometric
measurements from Stecklum, Melnikov \& Meusinger (2007).}
\end{figure}

\subsection{IRAS 06068-0641}

A possible FUor-type eruption from the infrared source 
IRAS 06068-0641 was discovered by the 
Catalina Real-time Transient Survey (CRTS) on 10 November 2009
(Wils et al. 2009). The object lies inside a dark nebula to 
the south of the Monoceros R2 association, and is likely related to it.
Figure 4 shows the optical light curve (November 2009 - March 2011) of 
IRAS 06068-0641 in $R$-band. After  a significant increase in the brightness
from at least late November 2009 ($R$ $\sim$ 13.4 mag) to January 2010 
($R$ $\sim$ 12.2 mag), a general decline in the source brightness can be 
seen in recent observations.

The FUor class of eruptive low-mass YSOs display outbursts of $\sim$ 4 mag 
or more that last 
for several decades. EXors show smaller outbursts ($\Delta$m $\sim$ 
2 - 3 mag) that last from a few months to a few years and may occur repeatedly 
(Herbig 1977; Bell et al. 1995; Hartmann 1998). 
For the more than a year that we followed the source IRAS 06068-0641 
(January 2010 - March 2011), the brightness decreased by more than 3.5 mag in 
$R$ band and it is probably returning to its pre-outburst state.
It is therefore possible 
that we witnessed EXor behaviour, since the source variability had about 
the correct amplitude (Reipurth \& Aspin 2004; Ojha et al. 2005, 2006). 
Further photometric observations of this object are required to 
understand and classify the outburst happening now in IRAS 06068-0641.

\begin{figure}
\centerline{\includegraphics[width=11cm]{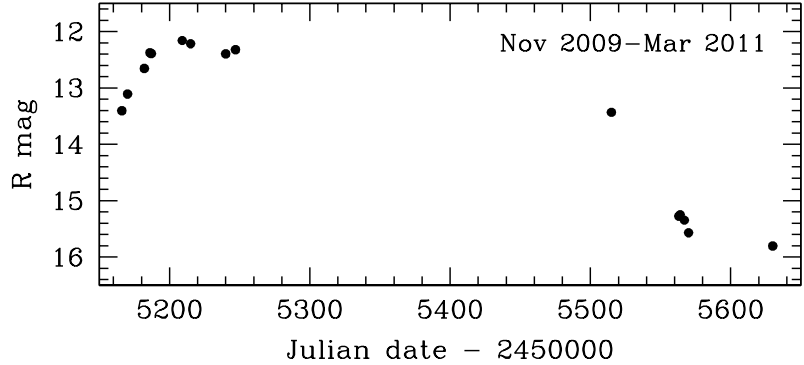}}
\vspace*{-5.5cm}
\caption{The optical light curve of IRAS 06068-0641 in $R$ band.}
\end{figure}

\section*{Acknowledgements}

The authors thank the staff of HCT operated by Indian Institute
of Astrophysics (Bangalore) and IGO operated by Inter-University Centre
for Astronomy \& Astrophysics (Pune) for their assistance and support during
observations. 



\begin{thebibliography}{}
\bibitem{}Aspin C., 2008, IAU Circ., 8969
\bibitem{}Bell K.~R., Lin D.~N.~C., Hartmann L.~W., Kenyon S.~J., 1995, ApJ, 444, 376
\bibitem{}Hartmann L., 1998, in Accretion Processes in Star Formation. Cambridge Univ. Press, Cambridge 
\bibitem{}Hartmann L., Hinkle K., Calvet N., 2004, ApJ, 609, 906
\bibitem{}Herbig G.~H., 1977, ApJ, 217, 693
\bibitem{}Herbig G.~H., Petrov P.~P., Duemmler R., 2003, ApJ, 595, 384
\bibitem{}Kastner J., Richmond M., Simon T., Grosso N., Weintraub D.,
Hamaguchi K., 2006, CBET, 760, 1
\bibitem{}Landolt A.~U., 1992, AJ, 104, 340
\bibitem{}Lynds B.~T., 1962, ApJS, 7, L1
\bibitem{}McNeil J.~W., 2004, IAU Circ., 8284
\bibitem{}Ojha D.~K., et al., 2005, PASJ, 57, 203
\bibitem{}Ojha D.~K., et al., 2006, MNRAS, 368, 825
\bibitem{}Ojha D.~K., Ghosh S.~K., Kaurav S.~S., Bhatt B.~C., 
Sahu D., Tej, A., 2008, IAU Circ., 9006
\bibitem{}Pawade V.~S., et al., 2010, ASInC, Vol. 1, pp 243-244, 
ed. D. K. Ojha
\bibitem{}Reipurth B., Aspin C., 2004, ApJ, 606, L119
\bibitem{}Reipurth B., Bally J., 2001, ARA\&A, 39, 403
\bibitem{}Samus N.~N., 2004, IAU Circ., 8354
\bibitem{}Sandell G., Weintraub D.~A., 2001, ApJS, 134, 115
\bibitem{}Stecklum B., 2006, CBET, 690, 1
\bibitem{}Stecklum B., Melnikov S.~Y., Meusinger H., 2007, A\&A, 463, 621
\bibitem{}Stetson P.~B., 1987, PASP, 99, 191
\bibitem{}Vittone A.~A., Errico L., 2005, MmSAI, 76, 320
\bibitem{}Wils P., Greaves J., Drake A.~J., Catelan M., 2009, CBET, 2033, 1
\end{thebibliography}
\end{document}